\pdfoutput=1
\documentclass[prd,nofootinbib,preprintnumbers,twocolumn,
showpacs,floatfix,superscriptaddress]{revtex4-1}

 \usepackage{auncial}
\usepackage{scrextend}
\usepackage{relsize}
\usepackage{amsmath}
\usepackage{amssymb}
\usepackage{epsfig}
\usepackage{graphicx}
\usepackage{hyperref}
\usepackage{dcolumn}   % needed for some tables
\usepackage{tabu}
\usepackage{boldline}
\usepackage{slashed}
\usepackage{multirow}
\usepackage{color}
\usepackage[normal]{subfigure}
\usepackage{rotating}
\usepackage[margin=0.9in,a4paper]{geometry}
\usepackage[table]{xcolor}
\usepackage{enumitem}
\usepackage[utf8]{inputenc}
\usepackage{colortbl}
\usepackage{array,multirow}
\definecolor{nicered}{rgb}{0.7,0.1,0.1}
\definecolor{nicegreen}{rgb}{0.1,0.5,0.1}
\definecolor{red}{rgb}{1.0, 0, 0}

\def\eq#1{{Eq.~(\ref{#1})}}
\def\eqs#1#2{{Eqs.~(\ref{#1})--(\ref{#2})}}

\def\Table#1{{Table~\ref{#1}}}

\def\vev#1{\left\langle #1\right\rangle}
\def\abs#1{\left| #1\right|}

\def\Tr{\mbox{Tr}\,}

\def\diag{\mbox{diag}\,}

\def\etc{\hbox{\it etc}{}}

\def\gsim{\raise0.3ex\hbox{$\;>$\kern-0.75em\raise-1.1ex\hbox{$\sim\;$}}}
\def\lsim{\raise0.3ex\hbox{$\;<$\kern-0.75em\raise-1.1ex\hbox{$\sim\;$}}}
\def\mb[#1]{\mathbf{#1}}
\renewcommand{\bar}{\overline}

\definecolor{LightCyan}{rgb}{0.88,1,1}
\definecolor{piggypink}{rgb}{0.99, 0.87, 0.9}
\definecolor{applegreen}{rgb}{0.55, 0.71, 0.0}
\definecolor{darkpastelgreen}{rgb}{0.01, 0.75, 0.24}
\definecolor{green-yellow}{rgb}{0.68, 1.0, 0.18}

\newcommand{\beq}{\begin{equation}}
\newcommand{\eeq}{\end{equation}}
\newcommand{\beqa}{\begin{eqnarray}}
\newcommand{\eeqa}{\end{eqnarray}}
\begin{document}
% ----------------- preprint numbers ------------------
% \begin{frontmatter}

% ------------- Title and authors ---------------------

\title{Gauge leptoquark as the origin of $B$-physics anomalies}

\author{Luca Di Luzio}
\email{luca.di-luzio@durham.ac.uk}
\affiliation{\normalsize \it 
Institute for Particle Physics Phenomenology, Department of Physics, Durham University, DH1 3LE, Durham, United Kingdom}
\author{Admir Greljo}
\email{admir@physik.uzh.ch}
\affiliation{\normalsize\it Physik-Institut, Universit\"{a}t Z\"{u}rich, CH-8057 Z\"{u}rich, Switzerland} 
\affiliation{\normalsize\it Faculty of Science, University of Sarajevo, Zmaja od Bosne 33-35,
71000 Sarajevo, Bosnia and Herzegovina}
\author{Marco Nardecchia}
\email{marco.nardecchia@cern.ch}
\affiliation{\normalsize\it Theoretical Physics Department, CERN, Geneva, Switzerland}

% ------------------------------------------------------
\begin{abstract}
  \noindent

The vector leptoquark representation, $U_\mu = (3,1,2/3)$, was recently identified as an exceptional single mediator model to address experimental hints on lepton flavour universality violation in semileptonic $B$-meson decays, both in neutral ($b \to s \mu \mu$) and charged ($b \to c \tau \nu$) current processes. Nonetheless, it is well-known that massive vectors crave an ultraviolet (UV) completion. We present the first full-fledged UV complete and calculable gauge model which incorporates this scenario while remaining in agreement with all other indirect flavour and electroweak precision measurements, as well as, direct searches at high-$p_T$. The model is based on a new non-abelian gauge group spontaneously broken at the TeV scale, and a specific flavour structure suppressing flavour violation in $\Delta F = 2$ processes while inducing sizeable semileptonic transitions.

\end{abstract}

\maketitle

\section{Introduction} The increasing set of experimental anomalies in semileptonic  $B$-meson decays might be the long-awaited signal of physics beyond the Standard Model (SM). That includes several measurements of lepton flavour universality (LFU) violation: $i)$ deviations from $\tau / \mu$ (and $\tau / e$) universality in $b \to c \ell \nu$ charged currents~\cite{Lees:2013uzd,Hirose:2016wfn,Aaij:2015yra}, and $ii)$ deviations from $\mu / e$ universality in $b \to s \ell \ell$ neutral currents~\cite{Aaij:2014ora}. The very recent LHCb measurements of $R^{\mu e}_{K^{(*)}}$~\cite{Aaij:2017vbb} and $R^{\tau \ell}_{D^{(*)}}$~\cite{LHCb:CERN} ratios reinforce the previous findings.  As emphasised in recent studies (see e.g.~\cite{Amhis:2016xyh,Capdevila:2017bsm,Altmannshofer:2017yso,Geng:2017svp,DAmico:2017mtc}), the overall statistical significance of the discrepancies in LFU measurements is at the level of $\sim 4\sigma$ for both charged and neutral current processes.  Further evidence of deviations from the SM predictions have been observed in the measurements of angular distributions of $B \to K^* \mu^+ \mu^-$ decay \cite{Aaij:2013qta,Aaij:2015oid}.

There have been several attempts in the literature towards a combined explanation of these anomalies (see e.g.~\cite{Buttazzo:2017ixm,Bhattacharya:2014wla,Alonso:2015sja,Greljo:2015mma,Calibbi:2015kma,Bauer:2015knc,Fajfer:2015ycq,Das:2016vkr,Boucenna:2016qad,Becirevic:2016yqi,Hiller:2016kry,Bhattacharya:2016mcc,Crivellin:2017zlb,Becirevic:2016oho,Cai:2017wry,Aloni:2017ixa,Dorsner:2017ufx,Barbieri:2015yvd,Buttazzo:2016kid,Barbieri:2016las}). Since the implied scale of new physics is rather low~\cite{DiLuzio:2017chi}, the main challenge is to reconcile it with the non-observation of related signals in (other) flavour changing processes (e.g.~\cite{Alonso:2016oyd}), electroweak precision observables, $\tau$ decays~\cite{Feruglio:2016gvd,Feruglio:2017rjo}, and high-$p_T$ searches~\cite{Faroughy:2016osc,Greljo:2017vvb}. Nonetheless, a coherent picture is emerging when invoking $i)$ a new dynamics in (mainly) left-handed semi-leptonic currents, and $ii)$ a flavour symmetry implying dominant couplings are to the third generation fermions~\cite{Buttazzo:2017ixm}.

A remarkably simple explanation of all the low-energy data is obtained by supplementing the SM with a single field -- vector leptoquark representation $U_\mu = (3,1,2/3)$ (see Fig.~(3) of Ref.~\cite{Buttazzo:2017ixm}). Importantly, leptoquarks~\cite{Dorsner:2016wpm} induce semi-leptonic transitions at tree level, while pure 4-quark and 4-lepton transitions arise only at one loop. However, the exceptional feature of this particular representation is the absence of tree-level down-quark-to-neutrino, as well as up-quark-to-charged-lepton transitions, naturally suppressing (a set of) otherwise strongly constrained observables.

In this paper, we show how to consistently embed the leptoquark $U_\mu$ into a non-abelian gauge theory spontaneously broken in the vicinity of the TeV scale, while still remaining consistent within a plethora of experimental constraints ranging from low-energy precision measurements to direct searches at the LHC.

\section{UV completion challenges}
Massive vectors require a UV completion in form of either a composite dynamics 
or a spontaneously broken gauge theory. The former approach was attempted for instance 
in Refs.~\cite{Barbieri:2015yvd,Buttazzo:2016kid,Barbieri:2016las}, 
where the vector leptoquark $U_\mu$ arises as a composite vector resonance of a new strong sector featuring an extended global symmetry, 
in analogy to composite Higgs or technicolor models. Such constructions, though plausible from the point of view of the naturalness problem of the electroweak scale, 
have the downside of not being fully calculable. For example, loop observables such as neutral meson mixing are quadratically divergent and can at most be  estimated via a hard cutoff regularization (see e.g.~\cite{Biggio:2016wyy}).   

Here we take a different approach by embedding the vector leptoquark $U_\mu$ into a spontaneously broken gauge theory. 
A clear option, suggested by the SM chiral content, is the Pati-Salam~(PS)~\cite{Pati:1974yy} model with gauge group 
$SU(4)_{PS} \times SU(2)_L \times SU(2)_R$.\footnote{The smaller subgroup $SU(4)_{PS} \times SU(2)_L \times U(1)_R$ 
would suffice for the scope of obtaining the leptoquark $U_\mu$.} 
However, a serious obstacle of such setup is the simultaneous presence of both left- and right-handed currents 
breaking lepton chirality, without being proportional to the corresponding lepton mass. 
Hence, the bounds from various LFV and FCNC processes 
push the mass of the leptoquark in the $100$ TeV ballpark 
\cite{Carpentier:2010ue,Kuznetsov:2012ai,Giudice:2014tma}. 
Allowing for a mixed embedding of the SM matter fields 
could help in suppressing right-handed currents in the down sector 
(e.g.~if $d_R \subset 6$ of $SU(4)_{PS}$). This, however, would 
still not be enough for $R_{D^{(*)}}$, due to the presence of a light $Z'$ from $SU(4)_{PS} \to SU(3)_c$ breaking 
with unsuppressed $\mathcal{O}(g_s)$ couplings to SM fermions \footnote{The resolution of both the 
$R_{D^{(*)}}$ and $R_{K^{(*)}}$ anomalies via a PS leptoquark $U_\mu$
was recently put forth in Ref.~\cite{Assad:2017iib}. In this respect, we reach a different conclusion.}. 

A crucial ingredient to circumvent the previous issues was recently proposed in Ref.~\cite{Georgi:2016xhm} 
in the context of a ``partial unification'' model in which 
the SM color and hypercharge are embedded into a $SU(3+N) \times SU(3)' \times U(1)'$ group.  
The latter resembles the embedding of color as the diagonal subgroup of two 
$SU(3)$ factors, as originally proposed in \cite{Hall:1985wz,Frampton:1987dn,Frampton:1987ut}. 
For $N=1$ one can basically obtain a massive leptoquark which 
does not couple to SM fermions, if the latter are $SU(3+N)$ singlets. 
A coupling of $U_\mu$ to left-handed SM fermions can be generated via the mixing 
with a vector-like fermion transforming non-trivially under $SU(4)' \times SU(2)_L$, 
as recently suggested in Appendix C of Ref.~\cite{Diaz:2017lit}. The latter model example, formulated in the context of leptoquark LHC phenomenology, is the starting point of our construction.
We go a step beyond and implement the necessary flavour structure to fit the $B$-anomalies, while keeping the model phenomenologically viable.

\section{Gauge leptoquark model} 
Let us consider the gauge group 
$G \equiv SU(4) \times SU(3)' \times SU(2)_L \times U(1)'$, 
and denote respectively by $H^\alpha_\mu, G'^a_\mu, W^i_\mu, B'_\mu$ the gauge fields, 
$g_4, g_3, g_2, g_1$ the gauge couplings and $T^\alpha, T^a, T^i, Y'$
the generators, with indices $\alpha = 1, \dots, 15$, $a = 1, \dots, 8$, $i=1,2,3$.  
The normalization of the generators in the fundamental representation is fixed by $\Tr T^\alpha T^\beta = \tfrac{1}{2} \delta^{\alpha\beta}$, \etc. 
The color and hypercharge factors of the SM gauge group $G_{\rm SM} \equiv SU(3)_c \times SU(2)_L \times U(1)_Y$ 
are embedded in the following way: 
$SU(3)_c = \left( SU(3)_4 \times SU(3)' \right)_{\rm diag}$ and $U(1)_Y = \left( U(1)_4 \times U(1)' \right)_{\rm diag}$, 
where $SU(3)_4 \times U(1)_4 \subset SU(4)$. 
In particular, 
$Y = \sqrt{\tfrac{2}{3}}T^{15} + Y'$, with $T^{15} = \tfrac{1}{2 \sqrt{6}} \text{diag}(1, 1, 1, -3)$. 

The spontaneous breaking $G \to G_\text{SM}$ happens via the scalar representations 
$\Omega_3 = \left( \bar 4, 3, 1, 1/6 \right)$ and $\Omega_1 = \left( \bar 4, 1, 1, -1/2 \right)$, 
which can be represented respectively as a $4 \times 3$ matrix and a $4$-vector 
transforming as $\Omega_3 \to U^*_4 \Omega_3 U_{3'}^T$ and $\Omega_1 \to U^*_4 \Omega_1$ under 
$SU(4) \times SU(3)'$. By means of a suitable scalar potential it is possible to achieve the following vacuum expectation 
value (vev) configurations \cite{inpreparation}
\beq  
\label{vevconf}
\vev{\Omega_3} = 
\left(
\begin{array}{ccc}
\tfrac{v_3}{\sqrt{2}} & 0 & 0 \\
0 & \tfrac{v_3}{\sqrt{2}} & 0 \\ 
0 & 0 & \tfrac{v_3}{\sqrt{2}} \\
0 & 0 & 0
\end{array}
\right) \, , \ \ 
\vev{\Omega_1} = 
\left(
\begin{array}{c}
0 \\ 
0 \\ 
0 \\
\tfrac{v_1}{\sqrt{2}}
\end{array}
\right) \, ,
\eeq
ensuring the proper $G \to G_\text{SM}$ breaking. 
Under $G_\text{SM}$ the scalar representations decompose as 
$\Omega_3 = (8,1,0) \oplus (1,1,0) \oplus (3,1,2/3)$ and 
$\Omega_1 = (\bar 3,1,-2/3) \oplus (1,1,0)$. 
After removing the linear combinations corresponding to the would-be Goldstone bosons, 
the scalar spectrum features a real color octet, two real and one pseudo-real SM singlets, 
a complex scalar transforming as $(3,1,2/3)$. 
The final breaking of $G_\text{SM}$ is obtained via the Higgs doublet field residing 
into $H = (1,1,2,1/2)$ of $G$ and acquiring a vev $\vev{H} = \frac{1}{\sqrt{2}} v$, with $v=246$ GeV. 

The gauge boson spectrum comprises three massive vector states belonging 
to $G / G_{\rm SM}$ and transforming as $U = (3,1,2/3)$, $g' = (8,1,0)$ and $Z' = (1,1,0)$ under $G_{\rm SM}$.  
From the scalar kinetic terms 
one obtains \cite{Diaz:2017lit,inpreparation}
\begin{align}
\label{MV}
M_{U} &= \tfrac{1}{2} g_4 \sqrt{v_1^2 + v_3^2} \, , \\
\label{Mgp}
M_{g'} &= \tfrac{1}{\sqrt{2}} \sqrt{g_4^2 + g_3^2} v_3 
\, , \\
\label{MZp}
M_{Z'} &= \tfrac{1}{2} \sqrt{\tfrac{3}{2}} \sqrt{g_4^2 + \tfrac{2}{3} g_1^2} \sqrt{v_1^2 + \tfrac{1}{3} v_3^2} 
\, .
\end{align}
Expressed in terms of the original gauge fields of the group $G$, the massive gauge bosons read 
\begin{align}
&U_\mu^{1,2,3} = \frac{1}{\sqrt{2}} \left( H^{9,11,13}_\mu - i H^{10,12,14}_\mu \right) \, , \\
&g'^a_\mu = \frac{g_4 H^a_\mu - g_3 G'^a_\mu}{\sqrt{g_4^2 + g_3^2}} \, , \ \ 
Z'_\mu = \frac{g_4 H^{15}_\mu - \sqrt{\frac{2}{3}} g_1 B'_\mu}{\sqrt{g_4^2 + \frac{2}{3} g_1^2}}\, , \nonumber
\end{align}
while the orthogonal combinations 
correspond to the massless $SU(3)_c \times U(1)_Y$ degrees of freedom of $G_{\rm SM}$ 
prior to electroweak symmetry breaking 
\begin{align}
g^a_\mu &= \frac{g_3 H^a_\mu + g_4 G'^a_\mu}{\sqrt{g_4^2 + g_3^2}} \, , \ \
B_\mu = \frac{\sqrt{\frac{2}{3}} g_1 H^{15}_\mu + g_4 B'_\mu}{\sqrt{g_4^2 + \frac{2}{3} g_1^2}} \, . \nonumber
\end{align}
The matching with the SM gauge couplings reads 
\beq 
\label{matchinggsgY}
g_s = \frac{g_4 g_3}{\sqrt{g_4^2 + g_3^2}} \, , \qquad g_Y = \frac{g_4 g_1}{\sqrt{g_4^2 + \frac{2}{3} g_1^2}} \, , 
\eeq
where $g_s = 1.02$ and $g_Y = 0.363$ are the values evolved within the SM up to the matching scale $\mu=2$ TeV. 
Since $g_{3,4} > g_s$ and $g_{4,1} > g_Y$, one has $g_{4,3} \gg g_1$. 
A typical benchmark 
is $g_4 = 3$, $g_3 = 1.08$ and $g_1 = 0.365$.

The would-be SM fermion fields (when neglecting the mixing discussed below), are charged under the $SU(3)' \times SU(2)_L \times U(1)'$ subgroup, but are singlets of $SU(4)$. Let us denote them as: $q'_L = (1,3,2,1/6)$, $u'_R = (1,3,1,2/3)$, 
$d'_R = (1,3,1,-1/3)$,  $\ell'_L = (1,1,2,-1/2)$, and $e'_R = (1,1,1,-1)$. These representations come in three copies of flavour. 
Being $SU(4)$ singlets, they do not couple with the vector leptoquark field directly.
To induce the required interaction, we add vector-like heavy fermions transforming non-trivially only under $SU(4) \times SU(2)_L$ subgroup. In particular, $\Psi_{L,R} = (Q'_{L,R}, L'_{L,R})^T = (4,1,2,0)$, where $Q'$ and $L'$ are decompositions under 
$SU(3)_4 \times U(1)_4 \subset SU(4)$. In order to address the $B$-physics anomalies, at least two copies of these representations are required.  
When fermion mixing is introduced (cf.~\eq{LYUK}) 
leptoquark couplings to SM fermions are generated. These are by construction mainly left-handed.
The field content of the model is summarized in \Table{fieldcontent}. 
\begin{table}[htp]
\begin{center}
\begin{tabular}{|c|c|c|c|c||c|c|}
\hline
Field & $SU(4)$ & $SU(3)'$ & $SU(2)_L$ & $U(1)'$ & $U(1)_{B'}$ & $U(1)_{L'}$ \\
\hline
$q'^i_L$ & 1 & 3 & 2 & $1/6$ & $1/3$ & 0 \\
$u'^i_R$ & 1 & 3 & 1 & $2/3$ & $1/3$ & 0 \\
$d'^i_R$ & 1 & 3 & 1 & $-1/3$ & $1/3$ & 0 \\
$\ell'^i_L$ & 1 & 1 & 2 & $-1/2$ & 0 & $1$ \\
$e'^i_R$ & 1 & 1 & 1 & $-1$ & 0 & $1$ \\ 
$\Psi^i_L$ & 4 & 1 & 2 & 0 & $1/4$ & $1/4$ \\
$\Psi^i_R$ & 4 & 1 & 2 & 0 & $1/4$ & $1/4$ \\
\hline
$H$ & 1 & 1 & 2 & 1/2 & 0 & 0 \\ 
$\Omega_3$ & $\bar 4$ & 3 & 1 & $1/6$ & $1/12$ & $-1/4$ \\ 
$\Omega_1$ & $\bar 4$ & 1 & 1 & $-1/2$ & $-1/4$ & $3/4$ \\ 
\hline
\end{tabular}
\end{center}
\caption{Field content of the model. The index $i=1,2,3$ runs over flavours, 
while $U(1)_{B'}$ and $U(1)_{L'}$ are accidental global symmetries 
(see text for further clarifications).   
}
\label{fieldcontent}
\end{table}

The full Lagrangian\footnote{We also include a $\left[ \Omega_3 \Omega_3 \Omega_3 \Omega_1 \right]_1$ 
term in the scalar potential which is required in order to avoid unwanted Goldstone bosons \cite{inpreparation}.} 
is invariant under the accidental global symmetries $U(1)_{B'}$ and $U(1)_{L'}$,   
whose action on the matter fields is displayed in the last two 
columns of \Table{fieldcontent}. 
The vevs of $\Omega_3$ and $\Omega_1$ break spontaneously 
both the gauge and the global symmetries,  
leaving unbroken two new global $U(1)$'s: 
$B = B'+\frac{1}{\sqrt{6}}T^{15}$ and $L = L'-\sqrt{\frac{3}{2}}T^{15}$, 
which for SM particles correspond respectively to ordinary baryon and lepton number. 
These symmetries protect proton stability and make neutrinos massless. 
Non-zero neutrino masses 
require an explicit breaking of $U(1)_{L'}$, e.g.~via a $d=5$ effective operator 
$\ell' \ell' H H / \Lambda$, where $\Lambda \gg v$ is some UV cutoff.  

The fermions' kinetic term leads to the following left-handed interactions
\begin{align}
&\mathcal{L}_L \supset \frac{g_4}{\sqrt{2}} \bar{Q}'_L \gamma^\mu L'_L \, U_\mu + \textrm{h.c.}~\nonumber \\
& +\frac{g_4 g_s}{g_3} \left( \bar{Q}'_L \gamma^\mu T^a Q'_L - \frac{g_3^2}{g_4^2} \,\bar{q}'_L \gamma^\mu T^a q'_L \right) g'^a_\mu ~\nonumber \\
&+\frac{1}{6} \frac{\sqrt{3} \,g_4 g_Y}{\sqrt{2} \,g_1} \left( \bar{Q}'_L \gamma^\mu Q'_L - \frac{2 g_1^2}{3 g_4^2} \,\bar{q}'_L \gamma^\mu q'_L \right) Z'_\mu ~\nonumber\\
&-\frac{1}{2} \frac{\sqrt{3} \,g_4 g_Y}{\sqrt{2} \,g_1} \left( \bar{L}'_L \gamma^\mu L'_L - \frac{2 g_1^2}{3 g_4^2} \,\bar{\ell}'_L \gamma^\mu \ell'_L \right) Z'_\mu~,\label{eq:coupL}
\end{align}
and right-handed interactions
{\footnotesize
\begin{align}
&\mathcal{L}_R \supset \frac{g_4}{\sqrt{2}} \bar{Q}'_R \gamma^\mu L'_R \, U_\mu + \textrm{h.c.}~\nonumber \\
& +\frac{g_4 g_s}{g_3} \left( \bar{Q}'_R \gamma^\mu T^a Q'_R - \frac{g_3^2}{g_4^2} \,\left(\bar{u}'_R \gamma^\mu T^a u'_R + \bar{d}'_R \gamma^\mu T^a d'_R \right) \right) g'^a_\mu ~\nonumber \\
&+\frac{1}{6} \frac{\sqrt{3} \,g_4 g_Y}{\sqrt{2} \,g_1} \left( \bar{Q}'_R \gamma^\mu Q'_R - \frac{4 g_1^2}{3 g_4^2} \, \left ( 2\, \bar{u}'_R \gamma^\mu u'_R - \bar{d}'_R \gamma^\mu d'_R \right) \right) Z'_\mu ~\nonumber\\
&-\frac{1}{2} \frac{\sqrt{3} \,g_4 g_Y}{\sqrt{2} \,g_1} \left( \bar{L}'_R \gamma^\mu L'_R - \frac{4 g_1^2}{3 g_4^2} \,\bar{e}'_R \gamma^\mu e'_R \right) Z'_\mu~.\label{eq:coupR}
\end{align}}

\section{Flavour structure}  
The Yukawa Lagrangian is
\begin{align}
\label{LYUK}
&\!\! \mathcal{L}_{Y} \supset  - \bar{q}'_L \,Y_d \, H d'_R - \bar{q}'_L \,Y_u \, \tilde H u'_R - \bar{\ell}'_L \, Y_e \, H e'_R~  \\
&- \bar {q}'_L \, \lambda_q \, \Omega_3^T \Psi_R - \bar {\ell}'_L \, \lambda_\ell \, \Omega_1^T \Psi_R - \bar \Psi_L \, M \, \Psi_R + \text{h.c.}~, \nonumber
\end{align}
where $\tilde H = i \sigma_2 H^*$. Also, $Y_d$, $Y_u$, and $Y_e$ are $3 \times 3$ flavour matrices, $\lambda_q$ and $\lambda_\ell$ are $3 \times n_\Psi$, while $M$ is $n_\Psi \times n_\Psi$ matrix where $n_\Psi$ is the number of $\Psi$ fields.

In absence of the Yukawa Lagrangian the global flavour symmetry of the model is $U(3)_{q'}\times U(3)_{u'}\times U(3)_{d'}\times U(3)_{\ell'}\times U(3)_{e'} \times U(n_\Psi)_{\Psi_L}\times U(n_\Psi)_{\Psi_R}$. Using the flavour group, one can  without loss of generality start with a basis in which: $M = M^{\textrm{diag}}\equiv \diag(M_1,...,M_{n_\Psi})$, $Y_{d} = Y_{d}^{\textrm{diag}}$, and $Y_{e} = Y_e^{\textrm{diag}}$ are diagonal matrices with non-negative real entries, while $Y_u = V^\dagger Y_u^{\textrm{diag}}$, where $V$ is a unitary matrix.

After symmetry breaking, the fermion mass matrices in this (interaction) basis are
{\footnotesize
\begin{align}
\mathcal{M}_{d}&=\left(\begin{array}{cc}
\frac{v}{\sqrt{2}}Y_{d}^{\textrm{diag}} & \frac{v_{3}}{\sqrt{2}}\lambda_{q}\\
0 & M^{\textrm{diag}} 
\end{array}\right),\;
\mathcal{M}_{e}=\left(\begin{array}{cc}
\frac{v}{\sqrt{2}}Y_{e}^{\textrm{diag}} & \frac{v_{1}}{\sqrt{2}}\lambda_{\ell}\\
0 & M^{\textrm{diag}}
\end{array}\right),\nonumber\\
\mathcal{M}_{u}&=\left(\begin{array}{cc}
\frac{v}{\sqrt{2}}V^\dagger Y_{u}^{\textrm{diag}} & \frac{v_{3}}{\sqrt{2}}\lambda_{q}\\
0 & M^{\textrm{diag}} 
\end{array}\right),\;
\mathcal{M}_{\nu}=\left(\begin{array}{cc}
0 & \frac{v_{1}}{\sqrt{2}}\lambda_{\ell}\\
0 & M^{\textrm{diag}}
\end{array}\right).\label{eq:masses}
\end{align}}
\!\!\! These are $3+n_\Psi$ dimensional square matrices which can be diagonalized by unitary rotations $U(3+n_\Psi)$. 
For example, $\mathcal{M}_{e} = U_{e_L} \mathcal{M}_{e}^{\textrm{diag}} U_{e_R}^\dagger$, where the mass eigenstate, $\psi_{e_{L}}\equiv (e_{L}, \mu_{L}, \tau_{L}, E_{L}^1,...,E_{L}^{n_\Psi})^T$, are given by  $\psi_{e_{L}} = U^\dagger_{e_{L}} \psi'_{e_{L}}$, and similarly for the right-handed components. 

The vector boson interactions with fermions in the mass basis are obtained after applying these unitary rotations to \eqs{eq:coupL}{eq:coupR}.
Our goal is to get the right structure of the vector leptoquark couplings for $B$-physics anomalies as in Ref.~\cite{Buttazzo:2017ixm}, while suppressing at the same time tree-level FCNC in the quark sector mediated by the $g'$ and $Z'$ exchange. 
In order to do this both in up- and down-quarks, one can impose the \emph{complete flavour alignment} condition
$\lambda^{ij}_q \propto M^{ij}$. However, this setup predicts large couplings to valence quarks and is challenged by direct searches at the LHC.

In this work, we minimally introduce two extra vector-like fermion representations $\Psi$ ({\bf $n_\Psi = 2$}). The pattern of flavour matrices $\lambda_q$ and $\lambda_\ell$ is such that no mixing with the first, small mixing with the second, and large mixing with the third generation is obtained. In addition, there is a \textit{flavour alignment} of the matrix $M$ with the quark mixing matrix $\lambda_q$. More precisely, in the basis of Eq.~\eqref{eq:masses}
\begin{equation}
\label{lamqmat}
\lambda_q = 
\begin{pmatrix} 
0 & 0\\
\lambda_q^{s} & 0\\
0 & \lambda_q^{b}
\end{pmatrix}~,
\end{equation}
{with $\abs{\lambda_q^{s}} \ll \abs{\lambda_q^{b}}$.}
The main implications of this setup are: $i)$ the absence of tree-level FCNC in the down-quark sector due to the $g'$ and $Z'$ exchange, 
and $ii)$ suppressed couplings to the valence quarks relaxing the high-$p_T$ constraints. While potentially large contributions to 
$D$--$\bar D$ oscillation phenomena are possible {via CKM mixing}, we show in the next section that the present constraints can be satisfied. Therefore, we pursue the second scenario in the rest of this paper.

From a flavour model building perspective, one can identify $d'_R$, $\Psi_L$, $\Psi_R$ as triplets of the same flavour group $U(3)_{d'} \equiv U(3)_{\Psi_L} \equiv U(3)_{\Psi_R}$. 
The matrix $M$ is then proportional to the identity, while $\lambda_q$ and $Y_d$ are proportional to the same spurion $(3, \bar 3)$ of $U(3)_q \times U(3)_{d'}$, and hence simultaneously diagonalizable. The phenomenology of this assumption is not far from the benchmark example considered below.

\section{Low-energy constraints} The main goal of this analysis is to find a working benchmark point (BP) which fits well the low-energy data. To this purpose we perform a numerical scan over the fundamental parameters in the Yukawa Lagrangian in Eq.~\eqref{LYUK}. Using numerical diagonalization, we first fix the known SM fermion masses, and then calculate the vector boson interactions in the fermion mass basis. While we observe no flavour changing $g'$ and $Z'$ interactions involving SM down quarks, the leptoquark couplings, being the product of both quark and lepton left-handed rotations matrices, can have the correct form {in order to fit the $B$-anomalies}. An example of a good BP is: $v_1=541$ GeV, $v_3=845$ GeV, $M_1=900$ GeV, $M_2=611$ GeV, $\lambda_{q}^{s}=-0.093$, $\lambda_{q}^{b}=2.0$, $\lambda_{\ell}^{21}=0.14$, $\lambda_{\ell}^{22}=-0.27$, $\lambda_{\ell}^{31}=2.3$, and $\lambda_{\ell}^{32}=2.1$. Fixing $g_4 = 3.2$, the vector bosons' spectrum for this BP is: $M_{Z'} = 1.4$~TeV, $M_{U} = 1.6$~TeV, and $M_{g'} = 2.0$~TeV.

We next calculate the contribution of the vector leptoquark to the relevant low-energy observables entering the fit in Ref.~\cite{Buttazzo:2017ixm}: $R_{D^{(*)}}^{\tau/\ell}$, $\Delta C_9^\mu =-\Delta C_{10}^\mu$, as well as (radiatively induced) corrections to $Z$ and $W$ couplings, 
$\delta g^Z_{\tau_L}$, $\delta g^Z_{\nu_\tau}$, and $ | g^W_\tau / g^W_\ell|$. Matching to the notation used there, we find $C_U = 0.022$, $\beta_{s \mu} = 0.006$, $\beta_{s \tau} = 0.053$, and $\beta_{b \mu} =-0.25$, which is a significant improvement with respect to the SM (see Fig.~(4) in~\cite{Buttazzo:2017ixm}). While the SM point has $\Delta \chi^2_{\rm SM} \simeq 43$ with respect to the best fit point of the four parameter fit, our BP has $\Delta \chi^2_{\rm BP} \simeq 8$. The tension in the charged current anomaly is reduced but not completely relaxed (our BP corresponds to $R_{D}/R_{D}^{\rm SM} = R_{D^{*}} / R_{D^*}^{\rm SM} \simeq 1.1$, 
to be confronted with the experimental combination $1.237 \pm 0.053$ from \cite{Buttazzo:2017ixm}), 
while the neutral current anomaly is perfectly fitted ($\Delta C_9^\mu =-\Delta C_{10}^\mu = -0.66$). 
Although this numerical example proves our claim, it would be instructive to perform a more detailed survey of the parameter space of the model~\cite{inpreparation}. 

As already pointed out, this leptoquark representation does not contribute significantly to $B \to K^{(*)}\nu\nu$ and $\tau \to 3 \mu$. 
The $Z'$ contributes to leptonic $\tau$ decays, in particular, $\tau \to \mu \nu \nu$ and $\tau \to 3 \mu$. However, for the BP these are small when compared with the present limits. On the other hand, the dominant contribution to the $D$--$\bar D$ mixing is due to $g'$ and is just above the limits (we find $\Lambda_R = 1.5$~PeV and $\Lambda_I = 3.2$~PeV for the BP which is to be compared with the limits in Ref.~\cite{Isidori:2013ez}).
Finally, $B_s$--$\bar B_s$ mixing is induced at one loop via box diagrams involving the vector leptoquark and heavy charged lepton partners. We have checked that the mixing amplitude is finite and well within the limits for the chosen BP. 

We conclude this section by noting that the mixing with the vector-like fermions modifies $W$ and $Z$ boson interactions leading to important constraints from CKM unitarity, $\Delta F = 2$, and electroweak precision measurements~\cite{Fajfer:2013wca}. 
Prior to electroweak symmetry breaking, the mixing Lagrangian in the quark sector is $\mathcal{L}_{\rm mix} \supset  - ( (v_3/\sqrt{2})\, \lambda_q \, \bar{q}'_L + M \, \bar{Q}'_L )\, Q'_R + \text{h.c.}$, when considering only two mixing states. One can identify the heavy state with mass $M_{Q}^2 = M^2 + (\lambda_{q} v_{3})^2/2$ as $Q_L = \cos\theta_q \, Q'_L + \sin\theta_q \, q'_L$, where 
$\tan \theta_{q} =  \lambda_{q} v_{3} / (\sqrt{2} M)$.  The orthogonal combination, massless prior to the electroweak symmetry breaking, defines instead the SM component, $q_L = -\sin\theta_q \, Q'_L + \cos\theta_q \, q'_L$.\footnote{For example, the third generation quark mixing angle (due to $\lambda^b_q$) for the BP is large, $\sin \theta_q \simeq 0.9$. That is, the third family is almost entirely aligned with $\Psi_L$, unlike the first two. } This form can easily  be matched to the analysis of Ref.~\cite{Fajfer:2013wca} and, in particular, the constraints derived there apply. In general, the mixing angles induced after electroweak symmetry breaking are typically suppressed by the light fermion masses such that $\tan\theta_{q} $ can easily be $\mathcal{O}(1)$ (see for example Fig.~5 in Ref.~\cite{Fajfer:2013wca} for the limits on $\sin \theta_{bD} \sim \frac{m_b}{m_D} \tan \theta_{q}$).

\section{Direct searches} The new states are subject to direct search constraints at the LHC. 
We briefly discuss the current bounds, starting from the leptoquark which sets the overall mass scale of the whole spectrum.

$\bullet$ $U$: The $R_{D^{(*)}}$ anomaly requires a leptoquark mass close to the TeV scale, 
in our benchmark $M_U = 1.6$ TeV. 
At the LHC, leptoquarks are pair produced via QCD interactions while their decays are fixed by the coupling strengths needed to fit the anomalies. This results in dominant decay modes of $U$ into 
quark and lepton doublets of the third generation. A bound on this configuration is obtained from a simple recasting \cite{DiLuzio:2017chi} of the CMS search \cite{Sirunyan:2017yrk} leading to $m_U > 1.3$ TeV for gauge leptoquarks.

$\bullet$ $Z'$: 
The peculiar gauge structure and matter embedding of the model implies suppressed $Z'$ couplings with the first generation fermions \cite{Diaz:2017lit}. 
For our BP $M_{Z'}=1.4$ TeV and the coupling strength to first generation quarks is 
$-\sqrt{2/3} \, g_Y g_1 / g_4 \, Y_{q_L,u_R,d_R} \simeq -0.034 \, Y_{q_L,u_R,d_R}$, 
where the latter denotes the SM hypercharge of quarks.
The $Z'$ is produced at the LHC through Drell-Yan processes mainly from valence quarks, 
and such small couplings allow to pass the stringent bounds from di-lepton searches 
involving either electrons and muons \cite{ATLAS:2017wce} or taus \cite{Aaboud:2017sjh} in the final state.

$\bullet$ $g'$: The coupling of $g'$ to first generation SM quarks 
is $- g_s g_3 / g_4 \simeq - 0.33$, while $m_{g'} = 2.0$ TeV for the BP.
The sizeable couplings and the relative lightness of $g'$ make LHC di-jets searches an important probe for the model.
However, bump searches loose in sensitivity when the width-to-mass ratio is too large. 
In particular, the interpretation of data is given up to decay widths of 15\% of the mass \cite{Aaboud:2017yvp}. 
For our BP the latter is naturally large ($\Gamma_{g'}/m_{g'} \simeq 26 \%$) thanks to a kinematically allowed 
extra decay channel into heavy quarks, and the interference effects are significant. 
A dedicated experimental analysis is therefore required to test this scenario.
On the other hand, limits on contact interactions from di-jet angular variables \cite{Alioli:2017jdo} turn out to be satisfied, due to the $g_3 / g_4$ suppression of the $g'$ couplings.

$\bullet$ Heavy fermions: the minimal setup features two generations of quark and lepton $SU(2)_L$ doublets, 
mixing with second and third generation SM fermions. 
Neglecting small electroweak symmetry breaking effects, for the BP we have: 
a $c/s$ partner with $m_{C/S} = 900$ GeV, a $b/t$ partner with $m_{B/T} = 1.3$ TeV, 
a $\mu/\nu_\mu$ partner with $m_{L_{\mu}} = 740$ GeV and a $\tau/\nu_\tau$ partner with $m_{L_\tau} = 1.4$ TeV. 
Third generation quark partners are heavy enough to comply 
with dedicated searches for bottom \cite{Aaboud:2017zfn} and top \cite{Sirunyan:2017ynj} partners. 
The BP also passes the limits from searches for lighter $c/s$ partners~\cite{Sirunyan:2017lzl} 
and $g'$-assisted production~\cite{Bai:2014xba} does not dominate over the QCD pair production.

$\bullet$ Heavy scalars: these comprise a new colored octet, a triplet and three SM singlets. Their mass is in the TeV ballpark, 
with a fine structure depending on the details of the scalar potential. 
However, they do not pose a particular phenomenological issue  
both from the point of view of direct searches (due to the reduced production cross-section) 
and indirect searches (since they couple to heavy--light fermions, flavour observables are naturally suppressed).

\section{Discussion and outlook} 

In this paper, we put forth a renormalizable UV completion of the vector leptoquark 
$U_\mu = (3,1,2/3)$, which was recently identified as an exceptional single-mediator model 
to address the combined explanation of $B$-anomalies both in neutral and charged currents.  
In short, the model is based on the gauge group $SU(4) \times SU(3)' \times SU(2)_L \times U(1)'$  
(which we creatively name as `4321' model) with a diagonal embedding of the SM color and hyperchage factors 
which ensures two important phenomenological features: $i)$ the leptoquark dominantly couples to left-handed 
currents via the mixing with vector-like fermions 
(as required by the anomalies and in order to suppress flavour constraints on the leptoquark mass) 
and $ii)$ the coupling of $g'$ ($Z'$) to first generation SM fermions is suppressed by a factor $g_3/g_4$ $(g_1/g_4)$, 
thus alleviating the constraints from direct searches. 

The large value of the $g_4$ coupling ($g_4 = 3.2$ in the BP), which is required by the phenomenological 
viability of the model, is admittedly
at the limit of perturbativity. 
Using the one-loop beta function criterium of \cite{Goertz:2015nkp}, namely $\abs{\beta_{g_4} / g_4} < 1$, 
we get $g_4 < 4 \pi / \sqrt{10} \simeq 4$.\footnote{Similarly, perturbative unitarity of leptoquark-mediated 
$2 \to 2$ fermion scattering amplitudes requires $g_4 < \sqrt{8 \pi} \simeq 5$ \cite{DiLuzio:2017chi}.}    
This notwithstanding, the theory can be consistently extended in the UV thanks to the asymptotic freedom 
of the $SU(4)$ factor. Note that the large value of $g_4$ at the onset of the renormalization group running 
helps in taming the emergence of UV Landau poles 
in the Yukawa couplings (which are also large for the BP). Eventually, however, the positive contribution from the Yukawas takes over 
and Landau poles can be generated (around $10^6$ GeV for the BP). A detailed analysis of the high-energy extrapolation of the model 
will be presented in \cite{inpreparation}.

A distinctive feature of the minimal model is the tight link among the gauge boson masses (cf.~\eqs{MV}{MZp}). 
In particular, $g'$ and $Z'$ cannot be arbitrarily decoupled from $U$, 
which is required to be around the TeV scale in order to explain the $R_{D^{(*)}}$ anomaly.   
One might ask whether extra sources of symmetry breaking contributing to the gauge boson masses 
can relax those tight relations. To this end, 
we have studied the contribution of all the one- and two-index tensor representations 
of $SU(4) \times SU(3)'$ to the gauge boson mass spectrum \cite{inpreparation}. 
The best option for simultaneously maximizing both the $g'$ and $Z'$ masses is a 
$(\overline{10},6)$, which yields $m_{g'}/m_U=\sqrt{2}$ and $m_{Z'}/m_U=1$, at the leading order in 
$g_4 \gg g_{3,1}$.\footnote{Other potentially useful representations are $(15,8)$ and $(\overline{10},1)$, 
yielding respectively $m_{g'}/m_U=3/2$ ($m_{Z'} = 0$) and $m_{Z'}/m_U=\sqrt{3}$ ($m_{g'}=0$).}   
Though some of the representations might help in rising the $g'$ and $Z'$ masses compared to $U$, 
none of them allows for a complete decoupling. We hence conclude that the presence of a light $Z'$ and $g'$, 
together with the leptoquark $U$, is a solid prediction and certainly provides 
a smoking-gun signature in high-$p_T$ searches. 

To sum up, we find the `4321' model particularly elegant for addressing the $B$-physics anomalies and 
the detailed phenomenological aspects will be investigated elsewhere~\cite{inpreparation}.

\section*{Note added}

After v1 of the present paper was posted to the arXiv there were a couple of developments 
related to our work, which we would like to comment on: 
\begin{itemize}
\item  
Ref.~\cite{Calibbi:2017qbu} considered a Pati-Salam gauge model with extra vector-like fermions.  
In their case the spectrum contains an extra $Z'$ (whose mass is related to the leptoquark by gauge symmetry breaking) 
with unsuppressed $\mathcal{O}(g_s)$ couplings to valence quarks. 
\item 
Version 2 of Ref.~\cite{Assad:2017iib} is now consistent with our statement that the Pati-Salam leptoquark is subject 
to stringent flavour constraints which rule out the explanation of the $B$-physics anomalies 
in the minimal Pati-Salam scenario.
\end{itemize}

\section*{Acknowledgments}
We would like to thank Gino Isidori for useful comments on the manuscript, and, in particular, for pointing out the possibility of a stronger $SU(3)$ flavour symmetry assumption.
We also thank Simone Alioli, Federico Mescia, Duccio Pappadopulo and Riccardo Torre for useful discussions. 
AG would like to thank David Marzocca and Andrea Pattori for useful discussions. This research was supported in part by the Swiss National Science Foundation (SNF) under contract 200021-159720.

 \bibliographystyle{apsrev4-1.bst}
  \bibliography{bibliography}

\end{document}